# Highly efficient optical filter based on vertically coupled Photonic crystal cavity and bus waveguide

Kapil Debnath,[1,*] Karl Welna,[1] Marcello Ferrera,[1] Kieran Deasy,[2] David G. Lidzey,[2] Liam O'Faolain[1]

[1]SUPA, School of Physics & Astronomy, University of St Andrews, North Haugh, St Andrews KY16 9SS, UK
[2]Department of Physics and Astronomy, University of Sheffield, Sheffield, S3 7RH, UK
*Corresponding author: kd343@st-andrews.ac.uk

We experimentally demonstrate a new optical filter design based on a vertically coupled photonic crystal cavity and a bus waveguide monolithically integrated on the silicon on insulator platform. The use of a vertically coupled waveguide gives flexibility in the choice of the waveguide material and dimensions, dramatically lowering the insertion loss while achieving very high coupling efficiencies to wavelength scale resonators.

Wavelength division multiplexing (WDM) has become an essential part of modern optical communication systems, maximizing the available bandwidth by simultaneously transmitting multiple channels through the same optical waveguide using multiple wavelengths. For different WDM functions such as optical filtering[1], modulation[2,3], multiplexing/demultiplexing[3], routing[4], optical waveguide-resonator systems have been studied extensively as due to their small foot print, they require very low operating power and promise chip scale integration.

Optical resonators can be classified into two categories[5] namely, travelling wave resonators such as the micro ring resonator and micro disk resonator, and standing wave resonators e.g. photonic crystal (PhC) cavities. Optical filters based on travelling wave resonators, where both the resonator and waveguide are monolithically fabricated on the same platform, have been the subject of a large volume of recent research. The primary reasons behind their popularity are their high extinction ratio and high quality factor. However, their Free Spectral Range (FSR) restricts the scalability of such devices in terms of the number of channels.

On the other hand, the design of PhC cavities with small mode number (i.e. single mode operation) and ultra-high quality factor is relatively mature[6,7]; nevertheless, due to their small size, efficient coupling of light in and out of the cavity is challenging making the overall extinction ratio of such device low. The conventional approach is to use a PhC line defect waveguide side coupled to a PhC cavity, but the line defect itself introduces additional Fourier components into the light cone of the cavity, which in turn reduces the overall extinction ratio[8].

In Ref. 1 a different optical filter design was proposed where a silicon bus waveguide is vertically coupled to a silicon PhC cavity. Here we develop insights into such vertical coupling between a waveguide and a PhC cavity, and show that very efficient transfer of light can be achieved between modes with dissimilar modal refractive indices. As a consequence, low refractive index waveguides with large mode areas may be used allowing very efficient interfacing with optical fibres. In addition, the monolithic vertically coupled geometry allows easy cascading of multiple cavities making it very promising for WDM applications.

Fig. 1. (a) Schematic diagram of the optical filter, where a bus waveguide is vertically coupled to a PhC cavity, b) cross-sectional

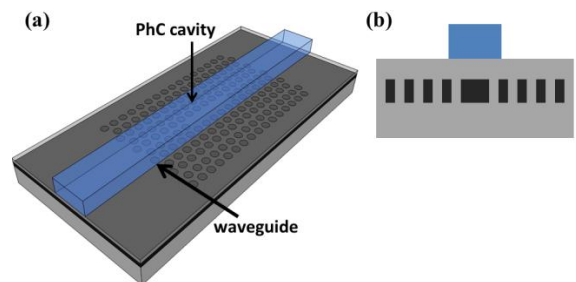

view of the cavity region (Color online)

Our vertical coupling scheme is shown in figure 1. A bus waveguide is placed vertically above a Silicon PhC cavity. An oxide barrier layer between the cavity and the waveguide works as a physical separation between the two optical modes. At the resonance wavelength of the cavity, light couples from the waveguide mode to the cavity mode. Under the weak coupling approximation, the transmission and reflection in the waveguide at resonance can be expressed as[5]:

$$T = \frac{Q_{total}^2}{Q_{cavity}^2}; \quad R = \frac{Q_{total}^2}{Q_{coupling}^2} \qquad (1)$$

Here, both the waveguide and the cavity are considered to be single mode within the wavelength range of interest. The cavity Q-factor, $Q_{cavity}$, is given by $1/Q_{cavity} = 1/Q_{design} + 1/Q_{fabrication}$, where $Q_{design}$ is given by the cavity design and $Q_{fabrication}$ by fabrication imperfections. $Q_{coupling}$ depends on the coupling efficiency between the bus waveguide and cavity mode. The overall Q-factor of the system $Q_{total}$ is:

$$\frac{1}{Q_{total}} = \frac{1}{Q_{cavity}} + \frac{1}{Q_{coupling}} \qquad (2)$$

It is clear from the above equations that in order to achieve large drop in transmission, i.e. $T \sim 0$, and large reflection, i.e. $R\sim1$ at resonance $Q_{design}$, $Q_{fabrication} \gg Q_{total}$. In other words, we need to increase the coupling efficiency while maintaining a high quality factor for the cavity ($Q_{cavity}$).

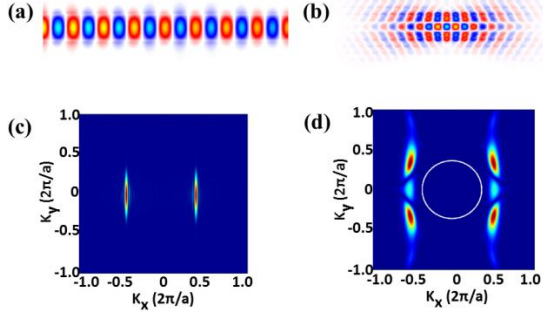

Fig. 2. Dominant electric field profile of (a) waveguide, (b) cavity, K-space distribution of the dominant field (c) waveguide, (d) cavity (Color online)

In-plane bus waveguide-cavity coupling configurations, which simplistically might be expected to affect only $Q_{coupling}$, actually have an impact on $Q_{cavity}$, disturbing the cavity structure and introducing additional K-space field components inside the light cone. 3D FDTD simulations have shown that placing an in-plane waveguide 3 rows away from an unmodified L3 cavity reduces $Q_{cavity}$ from 4982 to 2690. Hence, simultaneously achieving high coupling efficiency and maintaining high $Q_{cavity}$ becomes difficult. In contrast, when the waveguide is placed vertically, the presence of the waveguide does not significantly disturb the lattice structure of the PhC cavity and as a result a high $Q_{cavity}$ can be maintained.

The coupling efficiency depends mainly on two factors, a) the spatial overlap of the evanescent modes of the cavity and the waveguide and b) k-vector matching between the two modes. The spatial overlap between the two modes can be optimized simply by controlling the thickness of the barrier layer. Unlike in-plane coupling where the separation between waveguide and the cavity is a multiple of the lattice period, here we have complete freedom in choosing the separation. To ensure k-vector matching between the two optical modes, similar materials are used to guide both the optical modes in conventional coupled systems such as directional couplers. In the case of PhC cavities, the K-space distribution is expanded due to their ultra-small mode volume, helping achieve matching with the optical mode of a waveguide with dissimilar material. Figure 2 shows the real space and K-space distribution of the waveguide mode and the first order mode of the cavity (the design introduced in Ref. 7). Due to the rich K-space distribution of the cavity mode, a small change in the effective index of the waveguide can cause a dramatic change in the mode overlap and hence a large change in the transmittance of the whole system. Therefore, the coupling efficiency between the waveguide mode and the cavity mode, can be controlled either by changing the physical properties of the waveguide, such as the dimensions or material, the lattice structure of the PhC cavity or the thickness of the barrier layer. All these may be controlled with very high precision. Importantly, the coupling efficiency is not controlled by feature whose dimension is critical in terms of lithography, a significant advantage over many other approaches.

We fabricated the PhC cavity in the Silicon–on–insulator (SOI) platform[9]. Instead of the undercutting typically used, the sample was instead spun with a flowable oxide (FOx) containing hydrogen silsesquioxane (commercially available FOx-14 from Dow Corning) which fills the PhC holes. After curing the FOx has a refractive index similar to that of buried oxide (~1.4). The advantage of using FOx is threefold, firstly, it helps to achieve a symmetric cavity mode ensuring a high Q-factor[10], secondly it offers mechanical stability and, finally FOx layer acts as the buffer layer between the waveguide and cavity. A thickness of 200nm was used to achieve efficient coupling while keeping $Q_{total}$ reasonably high. To demonstrate the full versatility of our coupling technique, we have used two different materials for the top waveguide: ZEP 520A (n≈1.52) and $Si_3N_4$ (n≈1.88). The waveguides were patterned in the ZEP layer using e-beam lithography, giving a cross-section of 3×2um². For the $Si_3N_4$ waveguides, a 500nm layer of $Si_3N_4$ was deposited using Plasma Enhanced Chemical Vapour Deposition (PECVD) and waveguides were then patterned onto $Si_3N_4$ layer using e-beam lithography and dry etching.

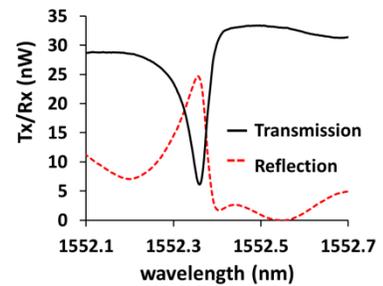

Fig. 3. Transmission and reflection spectra of a vertically coupled cavity waveguide system with a ZEP bus cross-section of 3×2um² (Color online). The extinction ratio is the ratio between the on and off resonance transmission

Figure 3 shows a typical transmission and reflection spectra (measured using endfire setup) of the coupled cavity–waveguide system where the waveguide material was ZEP 520A. At resonance, the extinction ratio is almost 7dB while the overall quality factor is maintained at 50000. From these results and using Eq.1 and Eq.2, we can infer that $Q_{coupling} \approx 60000$, and $Q_{cavity} \approx 110000$.

The ZEP-based waveguides were preliminarily used to demonstrate the low insertion loss (<3dB) of our optical filter. Subsequently, $Si_3N_4$ waveguides of varied width were used to demonstrate the effect of the K-space overlap on the coupling efficiency and thereby the overall extinction ratio of the filter and $Q_{cavity}$. This change in physical dimension varies the effective index and thereby moves the Fourier components of the waveguide mode along $K_x$ in K-space; as a result the overlap intensity with the cavity mode also changes. The width of the $Si_3N_4$ waveguides we used ranges from 0.8µm to 1.5µm. In most of this range the waveguides support a single quasi-TM mode. However, the higher order mode that occurs for width>1.2µm, accounts for a very little energy if compared with the fundamental mode and it does not have k-space overlap with the cavity modes. For this reasons, only the fundamental mode is taken into account in our analysis.

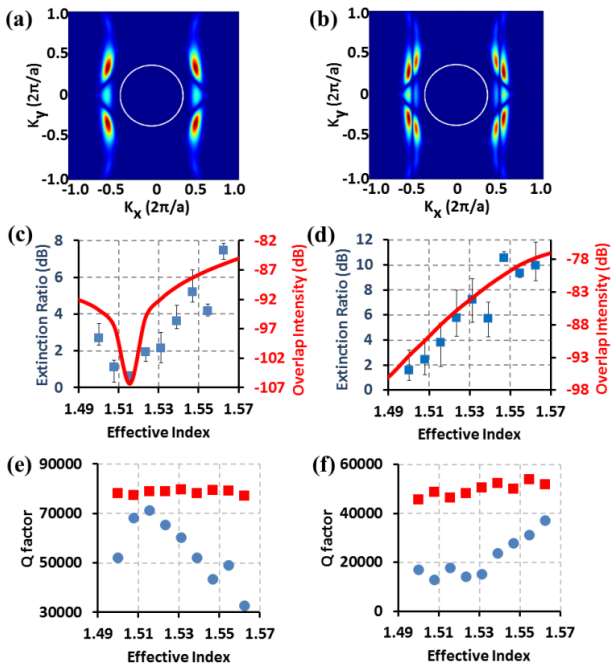

Fig. 4. K-space distribution of dominant electric field of (a) mode1, (b) mode2, Extinction ratio (blue squares) and k-space overlap intensity (red line) for different waveguide dimensions for (c) mode1, (d) mode2, $Q_{total}$ (red squares) and $Q_{cavity}$ (Blue circles) for different waveguide dimensions for (e) mode1, (f) mode2 (Color online)

The PhC cavity design used here is similar to that used in the previous experiment (figure 1). A design was chosen that exhibits two high Q modes so as to highlight the importance of the k-space overlap. Figure4 (a), (b) shows the k-space distribution of the first two modes of the cavity. In Figure4 (c), (d), the solid lines are the calculated K-space overlap intensities as the effective index of the waveguide is modified and the blue squares show the corresponding measured extinction ratios. Comparing the theoretical overlap intensities with experimental extinction ratios we note that both shapes match well. In particular, the dip above an effective index of 1.51 is present in both cases. In Figure4 (e), (f), blue circles represents $Q_{total}$ and red squares represents $Q_{cavity}$, which is calculated using equation (1). As one can see, due to different coupling efficiency the $Q_{total}$ is different for different waveguides, but $Q_{cavity}$ remains almost invariant to any change in coupling efficiency, thus providing very high extinction ratio, see eqn (1).

In this work an optical filter comprising of a vertically coupled cavity-waveguide system is proposed and experimentally demonstrated providing an efficient, ultra-compact filter. This vertical coupling technique allows complete flexibility in choosing the cavity and waveguide design and achieves high Q-factors and coupling efficiencies Furthermore, the use of low index waveguides ensures very low insertion loss for such devices. This is a very promising platform for the realization of key functions such as optical modulation[9] and filtering.

This work is supported by EPSRC under the UK silicon photonics project and through the EraNET Nano-Sci project LECSIN. Kieran Deasy and David Lidzey also acknowledge financial support from the Marie Curie ITN: ICARUS. The authors acknowledge Andrea di Falco for useful discussions.

### References
1. M. Qiu, Opt. Lett. 30, 1476 (2005)
2. T. Tanabe, K. Nishiguchi, E. Kuramochi, and M. Notomi, Opt. Express 17, 22505 (2009)
3. Q. Xu, B. Schmidt, J. Shakya and M. Lipson, Opt Express 14, 9430 (2006)
4. N. Sherwood-Droz, H. Wang, L. Chen, B. G. Lee, A. Biberman, K. Bergman, and M. Lipson, Opt. Express 16, 15915 (2008)
5. Y. Xu, Y. Li, R. K. Lee, and A. Yariv, Phys. Rev. E 62, 7389 (2000).
6. B.-S. Song, S. Noda, T. Asano, and Y. Akahane, Nature Materials 4, 207 (2005).
7. K. Welna, S. L. Portalupi, M. Galli, L. O'Faolain and T. F. Krauss, IEEE J. Quant. Electronics 48, 1177 (2012)
8. Z. Zhang and M. Qiu, Opt. Express 13, 2596 (2005)
9. Kapil Debnath, Liam O'Faolain, Frederic Y. Gardes, Andreas G. Steffan, Graham T. Reed, Thomas F. Krauss, Opt. Express, submitted.
10. S.-W. Jeon, J.-K. Han, B.-S. Song, and S. Noda, Opt. Express 18, 19361 (2010)